# Identification and Correction of Temporal and Spatial Distortions in Scanning Transmission Electron Microscopy

Kevin M. Roccapriore,[1], Nicole Creange, [1,2], Maxim Ziatdinov,[1,3,a] and Sergei V. Kalinin[1,b]

[1] Center for Nanophase Materials Sciences, Oak Ridge National Laboratory, Oak Ridge, TN 37831

[2] Materials Science and Engineering Department, North Carolina State University, Raleigh, NC 27606

[3] Computational Sciences and Engineering Division, Oak Ridge National Laboratory, Oak Ridge, TN 37831

Scanning transmission electron microscopy (STEM) has become the technique of choice for quantitative characterization of atomic structure of materials, where the minute displacements of atomic columns from high-symmetry positions can be used to map strain, polarization, octahedra tilts, and other physical and chemical order parameter fields. The latter can be used as inputs into mesoscopic and atomistic models, providing insight into the correlative relationships and generative physics of materials on the atomic level. However, these quantitative applications of STEM necessitate understanding the microscope induced image distortions and developing the pathways to compensate them both as part of a rapid calibration procedure for *in situ* imaging, and the post-experimental data analysis stage. Here, we explore the spatiotemporal structure of the microscopic distortions in STEM using multivariate analysis of the atomic trajectories in the image stacks. Based on the behavior of principal component analysis (PCA), we develop the Gaussian process (GP)-based regression method for quantification of the distortion function. The limitations of such an approach and possible strategies for implementation as a part of in-line data acquisition in STEM are discussed. The analysis workflow is summarized in a Jupyter notebook that can be used to retrace the analysis and analyze the reader's data.

---

[a] ziatdinovma@ornl.gov
[b] sergei2@ornl.gov

Scanning transmission electron microscopy (STEM) has emerged as the ultimate imaging tool for materials and nanostructures at the atomic and subatomic level[1–4]. The rapid development, commercialization, and adoption of aberration correctors[5,6] have enabled reliable atomically resolved imaging across multiple material classes ranging from semiconductors and oxides to 2D and quantum systems. Many other advances ranging from high-resolution electron energy loss spectroscopy (EELS) for single-atom chemical spectroscopy and mapping vibrational energy levels and phonons, vortex beams with orbital momentum for spin- and orbital imaging, 4D STEM and ptychography, and atomic manipulation have been reported[7–11].

However, among the many opportunities enabled by aberration correction, the one that stands on its own is quantitative structural imaging. Here, STEM imaging data is used to reconstruct positions of individual atomic columns (or individual atoms for 2D materials), ultimately transforming microscope-specific image data into materials-specific data (atomic coordinates). Using a combination of high-stability STEM and image post-processing[12–15], positional information with better than ~1 pm precision was reported[16].

This level of high-resolution structural imaging immediately presents fundamentally new opportunities for the exploration of the physics and chemistry of materials at the atomic level. A decade ago, work by Jia in TEM and Chisholm using STEM[17,18] pioneered mapping of polarization order parameters in ferroelectric materials, imaging modes that have since become mainstream[19–21]. Similarly, octahedral tilts and chemical strains have been reported for a variety of materials systems[22,23].

Beyond qualitative imaging, this structural data can be used to reconstruct the generative physics and chemistry of materials via matching with mesoscopic or atomistic models. The quantitative analysis of mesoscopically averaged order parameter fields derived from STEM data has been used to determine the nature of interface and gradient terms in the Ginzburg-Landau-type models of ferroelectrics[24]. This approach was further extended to quantify the flexoelectric interactions in perovskites, allowing for uncertainty quantification[25]. More recently, statistical distance minimization methods have been used to map the locally observed degrees of freedom of the Hamiltonians of lattice models, allowing for the reconstruction of the thermodynamics of solid solutions and interatomic forces. The use of Bayesian inference methods allows both the reconstruction of the generative model (e.g., interactions in the Ising case) and the uncertainty quantification on the model side.

However, the veracity of any such analysis also critically depends on the degrees of uncertainty in the STEM data, i.e., the presence of noise and systematic errors in the atomic coordinates derived from the STEM images. This in turn is comprised of two primary components; the relationship between the local STEM contrast and position of atomic nucleus and distortions in the image plane due to sample drift and fly-back delays as well as other non-idealities of the microscope's scanning system. For the former, the center of mass of the atomic column can be identified for the nucleus position with a high degree of precision[26–28]. This allows for quantitative analyses such as the octahedral tilts in the beam direction based on the column shape[29] once the low-order aberrations are excluded based on the calibration or presence of alternating order patterns within the sample. However, drift and distortion represent a more significant problem, as they are affected both by the mechanical stability of the microscope and the non-idealities of the electronic system. The latter represents the discrepancy of the ideal beam position imposed by the control voltages on the scanning coils and the real beam position on the sample plane.

We note that the drift and distortion parameters are strongly affected by specific microscope operating conditions, the day of the week, operation time, and settings of the electronics. Correspondingly, the time scales and stationarity of the distortion phenomena can differ significantly, necessitating a comprehensive analysis and dictating the choice of ameliorating strategies. Some of these will be stationary over long time periods (days and weeks) and thus predictable and in principle, correctable, and can include fly-back delays for large scan speeds. Some of these distortions can slowly change with time (hours), necessitating dynamic correction. Finally, some of the distortions can be rapidly changing with time, obviating correction strategies and for such dynamic distortions, reliable pathways to establish their presence, magnitude, and possible effects on the analysis of physical behaviors are required.

Here, we report the spatiotemporal analysis of STEM imaging based on multivariate analyses of atomic trajectories in dynamic STEM data (image stacks). This analysis allows separation of the long-term drift and random dynamic components of microscope distortions. Gaussian Process (GP) regression[30] is used to reconstruct the distortion of the image, allowing for distortion correction, and establish associated uncertainties in the reconstruction. We discuss the possible strategies for the selection of a reference image founded on materials-based (ideal lattice reconstruction) and image-based (imaging under optimal conditions) estimates, and generalization to other materials and instruments. The analysis workflow is summarized in a Jupyter notebook that can used to retrace the analysis and analyze the reader's data. The implications of this analysis on physics extraction and possible developmental strategies in STEM instrumentation and data analysis workflows are discussed.

**1. Image distortions in STEM**

The fundamental underpinning of the STEM is beam scanning, where the electron beam (e-beam) formed at the source via a set of control and aberration-correction lenses is scanned by the coils over the sample surface. The image is formed by detection of scalar intensity in high angle annular dark field (HAADF) operation, EELS, or a matrix of diffraction patterns (e.g., 4D STEM) while the e-beam moves over a fast scan axis and is subsequently shifted along the slow scan axis, rastering a specified region of the material's surface. The process can be continuous, or the e-beam can be shifted in a stepwise motion to enable grid-based detection. Notably, this scanning principle is similar to that in scanning probe microscopies, suggesting parallels between the two. It should also be noted that non-rectangular scan paths have been demonstrated, both with an a priori defined scan path or a scan path adapted based on image-based feedback[31].

However, this idealized process is associated by a number of non-idealities, deeply rooted in the nature of the scanning process. One of the most recognized issues during HAADF-STEM imaging is the fly-back delay, where the rapid change of the control bias during reversal of the beam motion at the end of slow scan line is smeared by the RC time constant of the control circuit, resulting in distortion of the image across the edge. Additionally, the beam motion induced by the scan coils can be non-orthogonal, which is a significant concern in determining the real atom positions when rotating the scan coils between samples. The relative parameters of a scan system can drift over time, e.g., due to changes in temperature and environmental conditions. Similarly, the relative position of the sample and the electron source can change over time. Superimposed on these electronic and mechanical drifts are environmental noises that often have complex non-white frequency characteristics.

Overall, these effects lead to distortion of the real sample plane $(x, y)$ compared to the scanned plane $(x', y')$. These distortions will have a complex spatiotemporal structure that can be described by the function $DF(x, y, t) = (x(t) - x'(t), y(t) - y'(t))$, i.e., the difference between ideal (i.e., intended by operator) and the real beam position. As is common in the analysis of inverse problems, the ideal beam position is unknown and unknowable (e.g., we do not have an independent way to find real atomic positions) and hence, practical analysis must rely on a number of simplifying assumptions of the process or material structure.

Here, we assume that the $DF(x, y, t)$ can be represented via the stationary part $DF_i(x, y)$ for each image, $i$, and the time-dependent part that describes the changes of $DF_i$ between images. This corresponds to coarse graining of the frequency spectrum of $DF(x, y, t)$ to the time scales of image acquisition. In this description, distortion will affect all images equally and is stationary. As such, it can be described by a single distortion function that can be algorithmically corrected either at the image analysis stage or on-the-fly during image acquisition. At the same time, if the distortion function changes with time the analysis becomes considerably more complicated and is demonstrated here. We defer analysis of the distortion function on the time scales below single image acquisition to further publications.

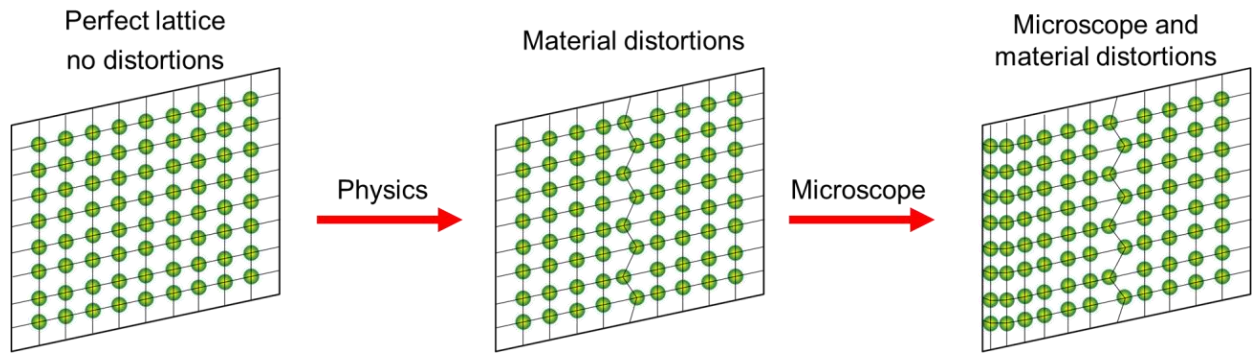

**Figure 1.** Sketch of non-distorted and otherwise perfect lattice that might be expected when imaging an atomic system (*left*); however, several physical processes govern that a material may undergo deviations from an ideal lattice site, e.g., presence of a zigzag reconstruction (*middle*). Microscope distortions are also present, e.g., fly-back, and therefore convolved with material distortions such that the final image is a combination of both physical distortions and microscope distortions (*right*).

It should also be noted that the detected image will contain superposition of the distortions induced by the imaging process and those present in the material, as illustrated in Figure 1. For example, for atomically resolved STEM images, the ferroelectric and ferroelastic orderings, strain fields, etc. will lead to displacements of atoms from ideal high-symmetry positions and in fact these distortions are often the target of the study. However, the *ad hoc* separation of the microscope and materials distortions is impossible and hence, the analysis must rely on the number of simplifying assumptions, including those on the length scale of phenomena of interest, the presence of an ideal periodic order, or the statistical properties of the system. For example, the difference in length scales of the distortion, e.g., between the unit-cell-level symmetry breaking due to the order parameters field and large-scale scan distortions can be used to separate the two. Alternatively, the time dependence of the image or behavior of the system with the translation of the image plane, where the image distortion shifts with the image plane where material's distortions remain fixed to a material can be used as a basis of the correction algorithm.

## 2. Experimental

Here, we aim to explore the spatiotemporal distortion dynamics in STEM imaging, including the presence of time-dependent components in the distortion function, estimate their magnitude, and develop the correction algorithm for the stationary part. As a model system, we select a $SrTiO_3$ (STO) single crystal that is expected to display minimal or no physical distortions; therefore, any distortions that are detected should be due to the microscope system itself. The STO sample is imaged in the (001) orientation using a NION UltraSTEM 200, $5^{th}$ order aberration-corrected STEM. The NION microscopes are well known for having ultra-stable stages, which can exhibit minimal specimen drift that is primarily a result of the magazine-based sample loading mechanism. Even so, drift is never fully eliminated and still must be considered, and therefore is a general concern for all STEM instruments. Imaging was performed at 200 kV with a convergence angle of 30 mrad and a probe current of roughly 20 pA. HAADF-STEM image stacks were acquired using several different scan parameters, e.g., fly-back delay time, beam lag time, pixel dwell time, number of pixels, as well as varying fields of view. To ensure consistency of the angle between the scan axes, the scan coil rotation angle is fixed at 0 degrees. We collected the image stacks for the purpose of analyzing the atom column trajectories as a function of time. We assumed that there were no beam-induced effects during the time scales we studied.

To gain insight into the spatial distortions incurred by the electronic scanning, we tested two extremes: short and long fly-back times. For short fly-back times, the distortion is strong enough to be visible even by eye. Changing to a long fly-back time certainly improves the fly-back distortion but can give a false sense of distortion-free imaging. Here, we develop a general scheme that can be applied for any image in a post-processing step, assuming that the microscope settings do not change. We showcase two common situations, one where there is strong distortion and one where it is weak. Emphasis is placed on the fact that this methodology is generalizable to other microscope systems given that the same assumption of unchanged microscope settings remains true.

## 3. Spatiotemporal trajectory analysis

We developed the workflow for analysis of the drift and spatiotemporal distortion in the STEM images based on the atomic trajectory analysis in the dynamic STEM data. This workflow is based on (a) the automatic atomic position extraction in each image frame in the STEM image stack, (b) reconstruction of atomic trajectories between the images, and (c) multivariate analysis of the trajectories. This workflow is based on previously reported deep learning tools augmented with the additional analytic functions.

The analysis starts from either the standard Digital Micrograph .dm3 or NumPy array image stack representing multiple observation of the system as an input. The .dm3 image is imported into the Python environment and converted to the NumPy object using the Python-adapted dm3 reader, pydm3reader[32]. Atomic positions are extracted using deep fully convolutional neural network with a U-net like architecture decorated with residual skip connections in both contracting and expanding paths[33,34]. The network's predictions are then used as initial guesses for standard Gaussian peak fitting of the atomic positions (Figure 2b, c).

To convert a set of atomic positions to trajectories, we utilized a k-d tree-based search[35] to identify the positions of a selected atomic column in all the movie frames in the presence of drift. To achieve this, we take the coordinates of the atomic column in the first frame and perform the k-d tree-based search for its nearest neighbor in the next frame. We then search for the nearest neighbor of the identified atomic column from the second frame in the third frame, and so on, until we reconstruct the entire trajectory. We repeat this procedure for all other atomic columns. This approach works as long as a drift-induced displacement between the two consecutive frames is less than a half of the atomic lattice distance. Note that in the absence of drift (or when the drift is negligibly small) we can find the entire trajectory at once, without a need to do the above described iterative "climbing," which significantly lowers the computation time. An example of several extracted trajectories is shown in Figure 2d.

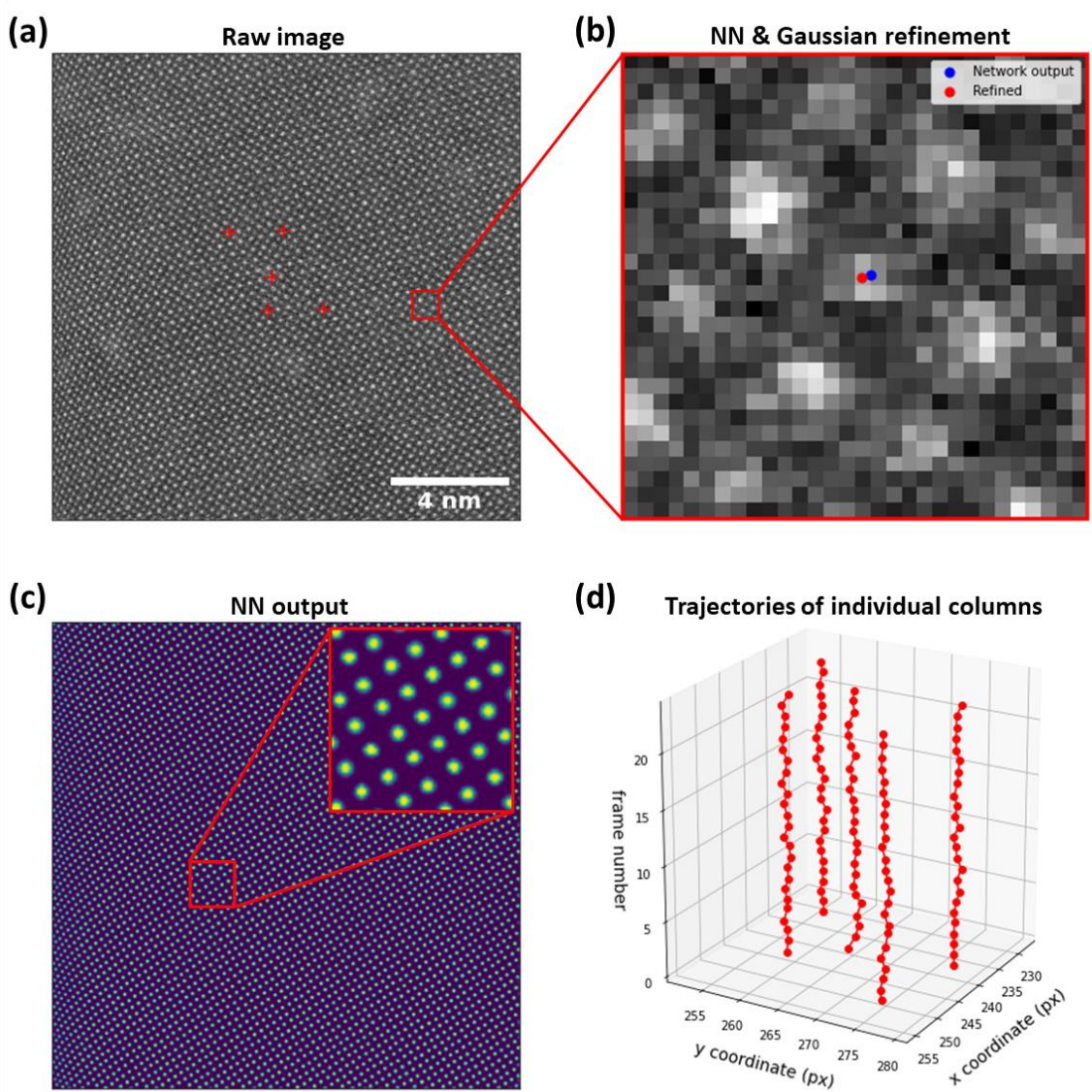

**Figure 2.** Extraction of atomic positions and trajectories for individual columns from image stack. (a) Raw experimental data representing a single (first) frame from image stack. (b) Positions of individual atoms are refined via standard Gaussian peak fitting using NN prediction as the initial guess. (c) Output of a fully convolutional neural network trained on simulated data. (d) Extracted trajectories of individual atomic columns (the *xy* "origin" for each column is shown by red cross in (a)).

With this analysis, the stack of STEM images is now converted into an array of atomic trajectories for each atom within the image plane, denoted as $T_{ij}(t) = (x_{ij}(t), y_{ij}(t))$, where $i, j$ define the lattice site and $x_{ij}$ and $y_{ij}$ are coordinates of the $(i,j)^{th}$ atom in the image plane. We note that without the loss of generality this analysis can be performed for non-crystalline or disordered materials, in which case the atoms can be simply numbered, $i, j \rightarrow k$. The analysis can be performed even if some of the atoms drift outside of the field of view during the experiment; however, given the difficulties of applying multivariate statistical methods to data on a non-uniform support, here we truncate the data set to the trajectories that can be observed across the full image stack.

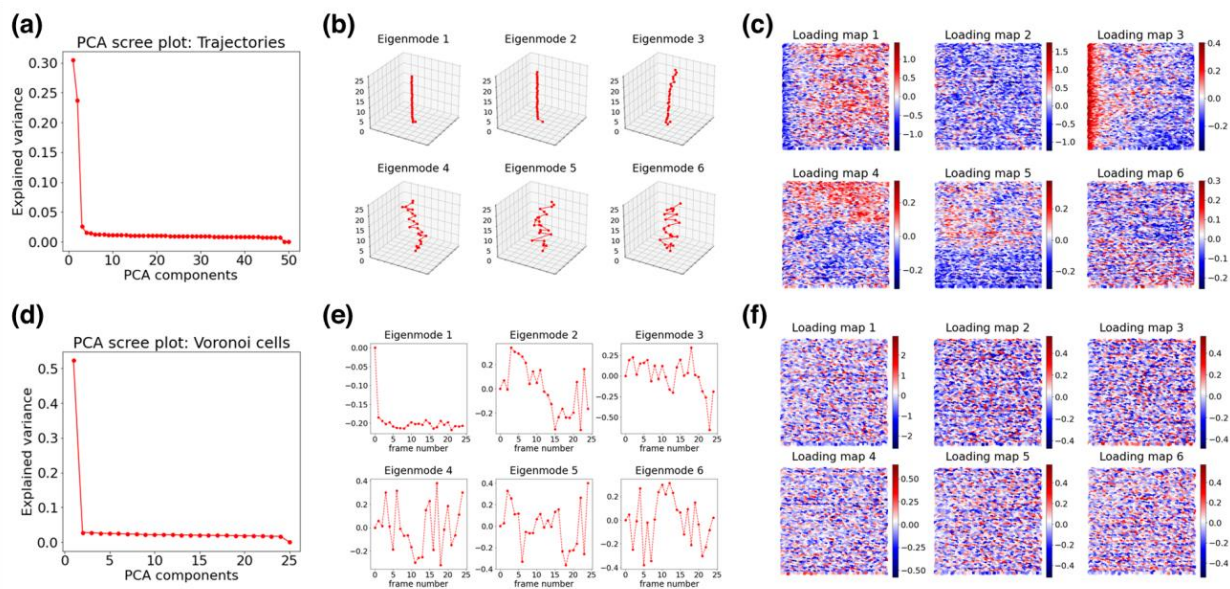

**Figure 3**. PCA analysis on trajectories (a-c) and Voronoi cell volumes (d-f) for a strong fly-back distorted image stack. Shown in each case are (a,d) scree plots, (b,e) eigenmodes, and (c,f) loading maps for the first six PCA components. Color scales in (c) and (f) are in units of pixels. Note the first frame in the stack appears to display anomalous behavior.

To gain insight into the spatiotemporal dynamics during microscope imaging, we analyze the array of the trajectories via unsupervised linear unmixing. We choose PCA as the simplest method that makes no specific assumptions about the structure of the distortion field and decompose the trajectories as

$$R(x, y, t) = \sum_{i=1}^{N} A_i(x, y) r_i(t) \qquad (1)$$

Here, $A_i(x, y)$ are the loading maps that represent the variability of the spectral behaviors across the composition space and $r_i(t)$ are the end members that determine the characteristic spectral behaviors. The number of components, *N*, is set at the beginning of the analysis and can be chosen based on the quality of decomposition, anticipated physics of the system, etc.

This choice of the multivariate method is predicated on the fact that there is no previously available information on the physical constraints on the trajectories, e.g., additivity, non-negativity, or sparsity. Therefore, the most general analysis method is used; however, other unmixing methods such as non-negative matrix factorization, independent component analysis, Gaussian mixture modeling, etc., can be used by using standard Python libraries if necessary and justified by additional knowledge on the physics of the imaging process.

The scree plot for the trajectory analysis of an image stack acquired using a short fly-back time is shown in Figure 3(a) where it is obvious that most of the observed dynamics are concentrated in the first three eigenmodes (PCA components). The structure of the eigenvectors and their spatial distributions are shown in Figure 3(b,c) and are rather remarkable. The first PCA component represents the average dynamics of the atoms and is equivalent to the frame-to-frame drift vector. The third (for this specific stack) PCA component is clearly dominated by transient behavior at the beginning of the image acquisition stack. This conclusion is further reinforced by analysis of the PCA loading maps shown in Figure 3(b). Here, the first and third PCA loading maps contain a discernible feature along the left edge of the image, consistent with the coupling between the noise and fly-back effects. The second component is more random, while the fourth and higher components show some form of random dynamics whereby the dynamics of these components contain clearly visible large-scale features with the feature size progressively decreasing for higher eigenmodes, behavior that is very different from that expected for a purely random process. These analyses clearly indicate that distortions present in the STEM images have clear systematic components. A dataset acquired using long flyback times was also demonstrated and the corresponding results shown in Figures 2 and 3 are depicted in Supplemental Figures S1 and S2.

We further explore the drift effect on the local materials descriptors that can be derived from the STEM image data, i.e., parameters such as unit cell volume, polarization, etc. Here, the nature of the model system allows only for the extraction of the unit cell volume. The corresponding scree plots are shown in Fig. 3(d) and several eigenvectors and loading maps are shown in Figures 3(e,f). Note that in this case the scree plot, eigenvectors, and loading maps are consistent with purely white noise-like behavior, where all the average information is concentrated in the first eigenmode whereas the remaining time dynamics are white noise processes.

**4. Gaussian process regression of drift function**

The analysis in Section 3 suggests that the image distortions in STEM have a complex spatiotemporal structure, naturally leading us to question whether the magnitude of these distortions can be ascertained and if they can be (partially) corrected. Both considerations are of obvious interest for quantitative STEM imaging and physics extraction since they constrain the level of materials-specific physical information that can be extracted from STEM image data.

We note that for a single image with an absence of any prior knowledge (i.e., Bayesian priors on distortion function), the distortions due to the imaging system and presence of materials-specific features are fundamentally inseparable. Correspondingly, the correction of image distortions should rely either on the presence of the sample with known or postulated ground truth (calibration standards) or on image acquisition with changing imaging parameters (e.g., a shift of the field of view), where stationary and non-stationary behaviors can be separated. Here, we adopt the approach based on a standard sample and defer more complex strategies for future studies.

The approach relies on the availability of a ground truth image, in which all the atomic coordinates are assumed to be known. Correspondingly, the scan distortions are defined with respect to this ground truth. The distortion function, $DF$, is defined as:

$$DF(x, y, k) = \{(x - x_{ideal}), (y - y_{ideal})\}(k) \tag{2}$$

This function is sampled at each lattice site, $i, j$, as the deviation of the observed atomic position from the ideal one. Here, $k$ is the index image frame representing the coarse-grained version of time, $t$. We note that $DF$ is a function of the position within the image plane, i.e., it can be defined for each pixel. Hence, establishing the microscope distortion is equivalent to the regression problem of determining the function $DF(x, y, k)$ from the experimental observations. Correspondingly, the time averaged $DF(x, y) = <DF(x, y, k)>$ represents the stationary distortion, whereas the time evolving part represents non-stationary behaviors. Stationary distortion can be corrected using a classical calibration whereas the time dependent part can be partially calibrated and quantified using the time-series analysis methods.

The natural question that arises in defining $DF(x, y, t)$ is choosing the ground truth image. This problem is common in Bayesian inference-based methods and multiple realizations of the ground truth image can be suggested depending on the specificity of the imaging mode. Some of the possible realizations include the point averages of the stack (after correction for linear drift), imaging performed with a much higher signal to noise ratio, a known calibration sample, etc.

We adopt a strategy based on the calibration sample that is assumed to have an ideal crystal lattice. In this case, several strategies to generate the ground truth image can be adopted, ultimately aiming to get the best (in the sense of a chosen optimization function) fit of lattice parameter and elementary unit cell. One such approach can be based on inverse Fourier filtering. In this approach, the image is Fourier transformed, the primary Fourier peaks are selected, and the Fourier density outside the peaks is set to zero using a bandpass filter. The inverse Fourier transform then gives an ideal version of the image. Note that these ground truth images can be created for each image in a stack and for the stack average, which balances the image noise and frame-to-frame drift. However, these Fourier-based methods are extremely sensitive to non-uniform distortions in the image and can be applied only if the magnitude of the distortion function, $DF(x, y)$, does not exceed half of the lattice parameter, i.e., $max\,[DF(x, y)] < a/2$ everywhere within the image. For the opposite case, $DF$ will be wrapped next to an atom, resulting in a characteristic ridge structure similar to phase wrapping in lock-in detection.

To avoid this problem, we adopt the approach based on indexing each atom in the image to the ideal lattice site $(i, j)$ and reconstruction of the ideal periodic lattice based on the indices and given point estimates for the lattice vectors. As such, this process comprises three steps; (a) constructing point estimates for lattice parameters, (b) lattice indexing, and (c) refinement of lattice estimates given the point estimates and indices. We note that an intrinsic feature of this analysis is the choice of a reference zero distortion point, in which the imaged and ideal lattice are matched exactly (chosen here to be the atom closest to the center of the scan). Supplemental Figure S3 includes details on the choice of a reference point.

To get point estimates for the lattice parameters, we analyzed the nearest neighbor distances in the central part of the image using the k-d tree-based search method. This works well for cubic systems where there is a single lattice spacing, but in the more general case a Fourier peak analysis approach can easily be used. The image is then Fourier transformed, and the primary Fourier peaks are selected and from these peaks the lattice rotation is calculated. Other transformations to obtain the rotation angle can be used, e.g., Radon transform. Note that a ground truth image is created for each image in a stack.

As a second step, we index each atom in the image to the associated lattice site $(i,j)$. The goal is to compare the $(i,j)^{th}$ atom in the ideal coordinate frame to the same $(i,j)^{th}$ atom in the experimental frame and compute differences in their positions, and repeat for all atomic columns. We note that this procedure is well-defined only in the absence of localized or extended structural defects, grains, or grain boundaries. In general, for point or extended defects the extended analysis can be based on corresponding lattice-level descriptions, e.g., introduction of a Burger's vector for dislocations. For systems with multiple crystalline regions, the problem transforms to dual indexing and segmentation. In this case, we avoid both of these complications via the use of a single crystal sample. Notably, the indexing process can be challenging, especially for heavily distorted images. We developed an algorithm that operates on analysis of the nearest neighborhoods of atomic columns. The details of the indexing process are found in the supplementary materials.

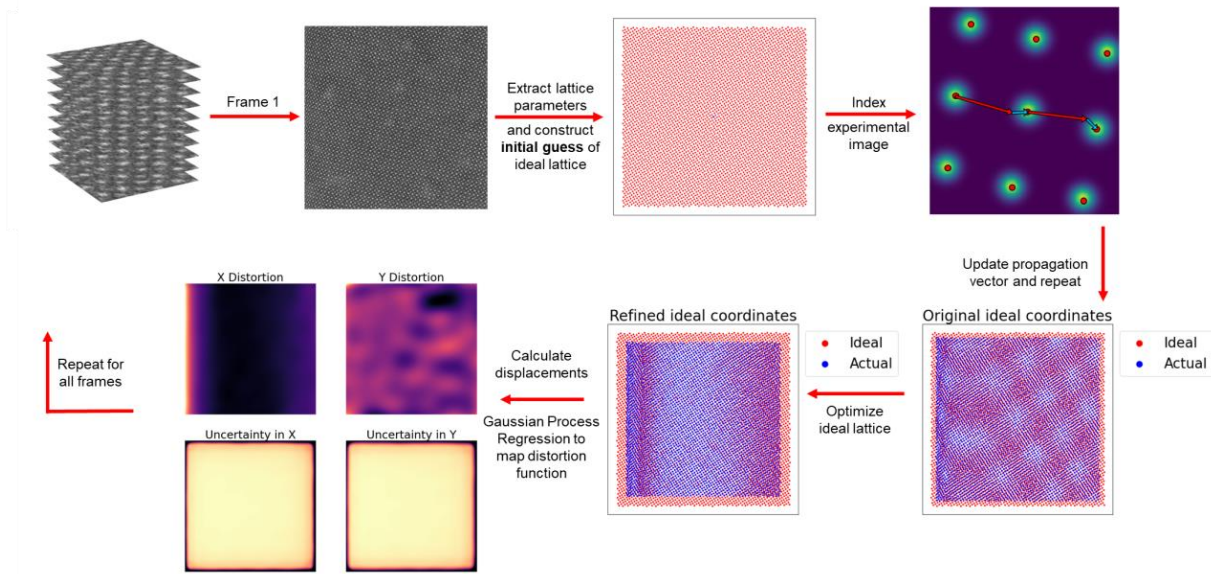

**Figure 4.** Process flow for obtaining distortion for each atom position in an atomically resolved STEM image. Note that the final GP regression step produces distortion function values and corresponding uncertainty at each pixel.

**Determination of single image distortion function**

With the ideal image defined, the problem of analyzing dynamic image distortions in a STEM image is then reduced for the regression problem for the $DF$ function. Here, we approach this problem using GP modeling[30]. In GP, a measured value, $S(\mathbf{x})$, is assumed to represent the function of interest, $S_0(\mathbf{x})$, with the addition of measurement noise, $N$. The noise is assumed to have zero mean, $E(N) = 0$, and to be Gaussian distributed (albeit this requirement can be relaxed and the noise structure can be a model parameter).

The measurements, $S(\mathbf{x}_i)$, for several values of the argument, $\mathbf{x}_i$, can then be considered to be a sample from the multivariate Gaussian distribution. The basic premise of the GP method is that the underlying Gaussian distributions are connected via the kernel function that defines the relationship between the corresponding covariances. For many practical problems, the kernel functions are translation invariant, i.e., $K(x_i, x_j) = K(x_i - x_j)$. The GP algorithm utilizes the information for several values of the argument and the kernel function to predict the properties in the unexplored location, i.e., acts as a universal interpolator. The key advantage of the GP method is that it yields both the function value and expected uncertainty that can be used for experimental planning and other workflows. It is also important to note that modern GP methods constrain the function form of the kernel function, but the corresponding hyperparameters (length scale of the kernel function, noise level, etc.) are determined self-consistently as a part of the regression process. In cases when the physics of the system are known, the kernel function and parameters can be constrained.

The GP approach for a scalar function of a scalar argument is illustrated in Figure 5. Here, the values of the function measured at several locations are shown and represent the ideal function and noise for strong fly-back distortion (a-c) and weak fly-back distortion (d-f). The results of the GP regression and associated uncertainty intervals are shown in terms of the X and Y components of the distortion. We implement the GP process using the *Pyro* library[37] in Python. The $DF$ function is defined following Eq. (2) where the ideal image is recalculated for each image in the stack. The GP regression is performed using the radial basis function (RBF) as a kernel. The uncertainty in the $DF$ can be found in Supplemental Material Figure S4 and is related to the density of sampling points, i.e., the number of atoms in the image. The uncertainty is significantly higher for the long fly-back dataset compared to the short fly-back dataset primarily due to the lower density of sampling points. The trade-off for increasing the density of atoms, however, is the fewer number of pixels per atom and hence, identifying precise atom coordinates becomes more difficult.

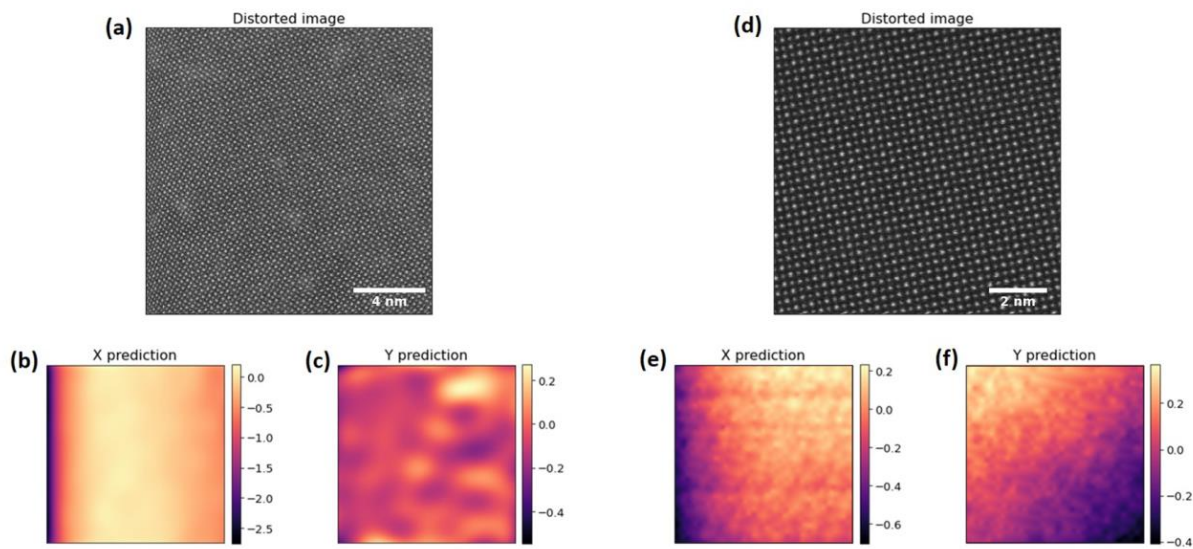

**Figure 5.** GP reconstruction of the *DF* for short fly-back (a-c) and long fly-back (d-f) times. Note the short fly-back effect is visualized on the left side of (a) and is clearly obtained in the X component of the distortion (b). Distortion map scales (b, c, e, and f) are in units of lattice spacings.

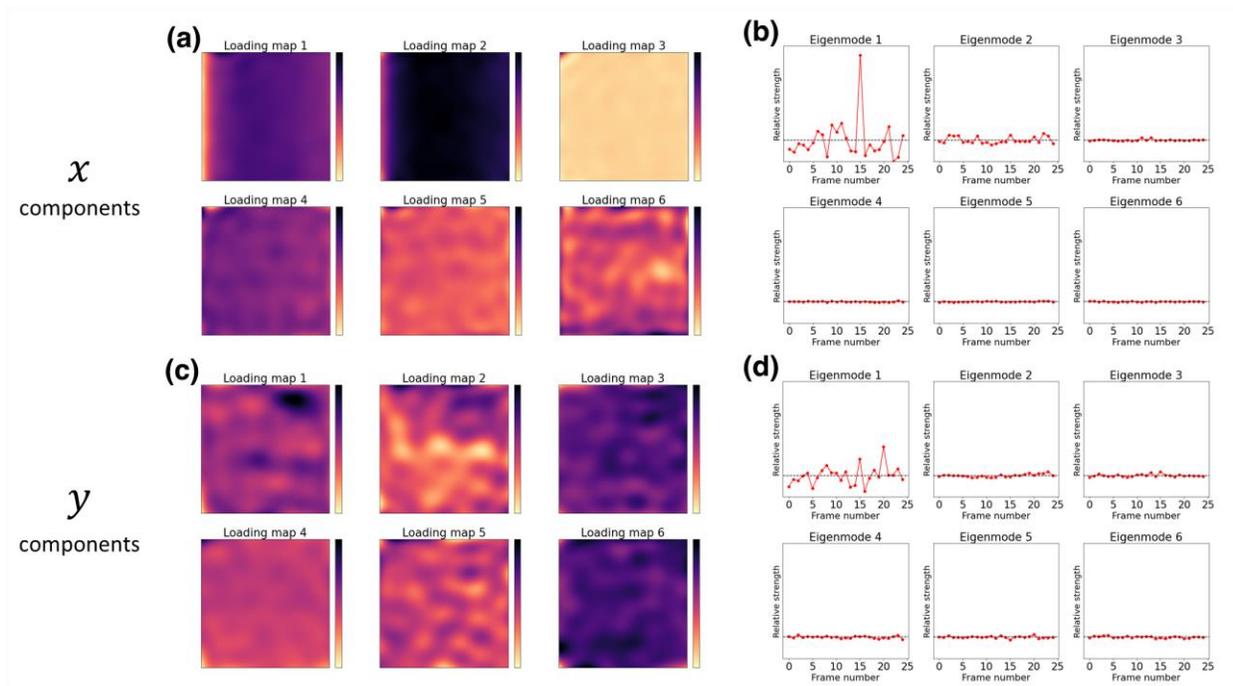

**Figure 6**. Multivariate analysis on the $DF(x,t)$ and $DF(y,t)$. First six separated components via PCA are shown. (a) and (c) depict loading maps of extracted components while (b) and (d) show corresponding eigenmodes as a function of coarse time (frame number)

Once the distortion maps are obtained for the stack of images representing a coarse-grained version of time, one can begin to analyze the complexity and time-stability of the $DF$. Using a common multivariate analytical approach, PCA is applied to the stack of each spatial component. Figure 6 depicts the first six PCA components for each axis in time. In this figure, data with strong fly-back distortion was used and as expected, is the most significant component in the horizontal axis. Note that the meaning of the loading maps observed indicate the components of distortion that are most significantly constant in time. We expect the fly-back to be constant over time and is the strongest distortion; therefore, it is seen as the first significant component. Of particular interest, however, is the fact that there is additional distortion present in the horizontal axis after the obvious fly-back is considered, specifically shown in components 4, 5 and 6 in Figure 6 (a,b). While they are understandably weaker distortions, they nonetheless persist. A similar case occurs along the vertical axis, where PCA component 1 persists over time more significantly than any other component.

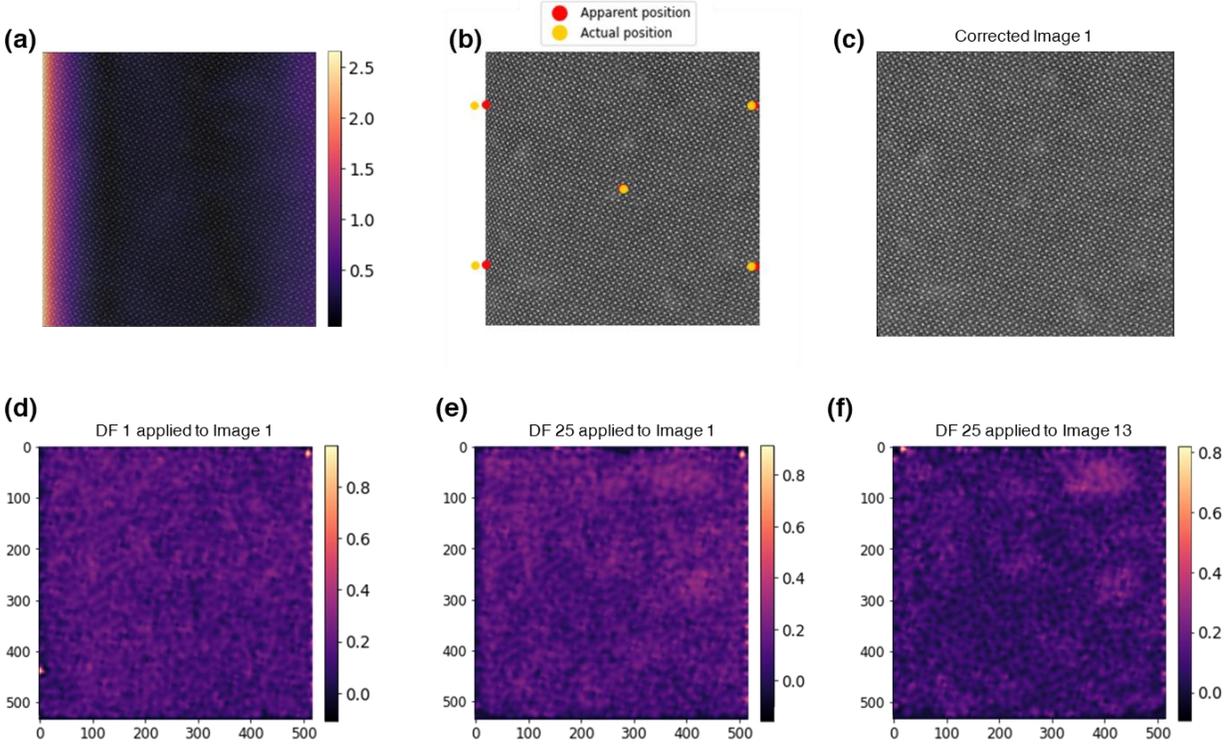

**Figure 7**. Distortion correction. (a) Original distorted image with scalar distortion overlay, (b) apparent and actual probe positions, (c) corrected image where majority of correction occurs in left portion (illustrated in (a)). Panels (d-f) show performance of distortion correction by correcting an image frame using a selected distortion function, then measuring atom column positions of corrected image relative to an ideal lattice, as was done to obtain the distortion functions. Note (d-f) depict the magnitude of distortion. All color bars are in units of lattice spacings.

**Distortion Correction**

Having learned the distortion maps for each image, one can then correct the distorted image, which is done by first resizing the distortion maps, if necessary, to be the same size as the input image. The distortion map then serves as a set of instructions for how much to shift each pixel in the original image. This is unlike typical distortion correction schemes that either make use of a known *DF* or a geometric transformation. Since the *DF* does not necessarily behave analytically, these routes are not possible. We employ a simpler pixel shift method because the distortion is well-defined at each pixel. This is justified because its length scale is significantly larger than the atomic spacing. Note that due to asymmetry in the horizontal and vertical distortions, the corrected image is usually not square. Figure 7 shows an overview of the correction process.

To complement the analysis over time, we apply the $DF$ learned from one image frame to a different image frame in the stack. For example, the first frame in the stack is corrected using the $DF$ learned from the final frame (Figure 7 (e)) and the middle frame in the stack is corrected using the same final frame distortion (Figure 7 (f)). It is clear that while the correction performs well overall and most of the distortion (>90%) is corrected, a complicated time evolution exists. To most accurately correct an image, the $DF$ from the same frame should be used, as shown in Figure 7 (c) and 7 (d). The fluctuations in the apparent distortion maps of the corrected images is likely a result of the atom identification process since specifying a level of uncertainty in atom fitting is not well-defined. Future endeavors may employ a modified atomic coordinate retrieval process, which can specify a degree of uncertainty. In the current state, using this approach with a given set of microscope parameters, one may learn the spatial distortions from an ideal crystal, then exchange to a different non-ideal crystal and apply the spatial distortion in order to extract quantitative information from a non-ideal system. Ideally the entire process should occur sequentially and in the same session, as the learned spatial distortion may change over time or when electronic parameters are adjusted.

Given that these experiments were carried out in an extremely stable microscope system, a natural question arises to whether such a stable system is crucial to the success of this methodology. We emphasize that our approach is quite generalizable to essentially any scanning microscope system that can produce sequential images that do not significantly drift from initial to final frame (i.e., majority of the same atomic columns remains in all frames). Different environments can produce different degrees of noise that can affect imaging in a variety of ways, however these types of external noise typically affect all images in a stack in the same way and therefore can be separated *post-hoc* by a variety of ML algorithms as was performed here with PCA.

To summarize, we have explored the spatiotemporal structure of the microscopic distortions in STEM imaging using multivariate analysis of the atomic trajectories in image stacks. The PCA analysis of the atomic trajectories shows complicated time dynamics containing random and average components. This analysis provides insight into the absolute stability of the imaging system.

We further extend this analysis to analyze the spatial distortions in the system. In this case, an ideal reference lattice is defined assuming perfect, discrete translational symmetry of the image and an approach for refinement of the lattice is proposed. The $DF$ can then be defined for each lattice site as the deviation of the observed atomic position from ideal. We further develop the GP-based regression method for the quantification of the $DF$ at all image pixels and quantify the associated uncertainties. The time dynamics of the $DF$ provide estimates of the stationary and non-stationary distortions in the STEM images. We note that this correction approach relies on a postulated ideal image and possible definitions based on low-distortion or ideally crystallographic ordered images are discussed. The analysis workflow is summarized in a Jupyter notebook that can be used to retrace the analysis and analyze the reader's data.

**Acknowledgement**

This work is based upon work supported by the U.S. Department of Energy (DOE), Office of Science, Basic Energy Sciences (BES), Materials Sciences and Engineering Division (K.M.R. and S.V.K.) and was performed and partially supported (MZ) at Oak Ridge National Laboratory's Center for Nanophase Materials Sciences (CNMS), a U.S. Department of Energy, Office of Science User Facility. Dr. Matthew Chisholm is gratefully acknowledged for providing the images used in this work, and he and Prof. Dr. Gerd Duscher (UTK) are acknowledged for multiple inspiring discussions.

12. Ophus, C., Ciston, J. & Nelson, C. T. Correcting nonlinear drift distortion of scanning probe and scanning transmission electron microscopies from image pairs with orthogonal scan directions. *Ultramicroscopy* **162**, 1–9 (2016).

13. Sang, X. & LeBeau, J. M. Revolving scanning transmission electron microscopy: Correcting sample drift distortion without prior knowledge. *Ultramicroscopy* **138**, 28–35 (2014).

14. Jones, L. *et al.* Smart Align—a new tool for robust non-rigid registration of scanning microscope data. *Adv. Struct. Chem. Imaging* **1**, 8 (2015).

15. Jones, L. & Nellist, P. D. Identifying and Correcting Scan Noise and Drift in the Scanning Transmission Electron Microscope. *Microsc. Microanal.* **19**, 1050–1060 (2013).

16. Yankovich, A. B. *et al.* Picometre-precision analysis of scanning transmission electron microscopy images of platinum nanocatalysts. *Nat. Commun.* **5**, 4155 (2014).

17. Jia, C.-L. *et al.* Unit-cell scale mapping of ferroelectricity and tetragonality in epitaxial ultrathin ferroelectric films. *Nat. Mater.* **6**, 64–69 (2007).

18. Chisholm, M. F., Luo, W., Oxley, M. P., Pantelides, S. T. & Lee, H. N. Atomic-Scale Compensation Phenomena at Polar Interfaces. *Phys. Rev. Lett.* **105**, 197602 (2010).

19. Lee, H. N., Christen, H. M., Chisholm, M. F., Rouleau, C. M. & Lowndes, D. H. Strong polarization enhancement in asymmetric three-component ferroelectric superlattices. *Nature* **433**, 395–399 (2005).

20. Nelson, C. T. *et al.* Domain Dynamics During Ferroelectric Switching. *Science* **334**, 968–971 (2011).

21. Yadav, A. K. *et al.* Observation of polar vortices in oxide superlattices. *Nature* **530**, 198–201 (2016).
23

**Supplemental Information**

Codes

The code used in this work is available as an executable Jupyter notebook at
https://git.io/JUWbS

Atomic indexing procedure

First, a separate image consisting of 2D Gaussian blobs centered at the coordinates given by Gaussian refinement of the neural network prediction is acquired such that we remove noise from the indexing process. . A starting 'seed' position for indexing is chosen at the center of the image as the center of the image is assumed to be the least distorted, where simplicity and symmetry are most ideal. The closest atom column to the 'seed' position and set as the [0,0] point of indexing. The local lattice vectors are calculated using the k-d tree-based search for a pre-defined number of nearest neighbors, which is set at 4 for STO. Basis vectors are identified by clustering the angle and magnitude of the local lattice vectors such that two degenerate vectors are not chosen as basis vectors. The calculated basis vectors, $v_1$ and $v_2$, are used to propagate lattice points and assigns an index to each atom column. Lattice propagation starts along iterative steps of $av_1$, where $a$ is an integer, before moving on to the next column and iterative again with steps of $av_1 + v_2$. In cases of rapid onset distortion, the atom closest to the proposed lattice site may not be the intended target, e.g., the lattice vector may have been over- or undershot and the correct atom is passed over or not reached. To account for this small distortion, upon each iterative step, the closest atom column to the proposed lattice site, the target atom, is found via a k-d tree-based search. Once the target atom is identified, the lattice vector, basis vector plus correction vector, is updated to accommodate deviations from the previous lattice vector and is propagated to find the next lattice site. In addition, each iterative step saves the proposed lattice site, the correction vector, $[i,j]$ index, and the corresponding atom column in an easily accessible data-frame. This process repeats until all atoms are found and indexed. This approach allows for highly distorted images, such as those with well-known fly-back compressions, to be indexed due to the simple fact that the lattice vector is constantly being updated based on previous lattice propagations.

To construct a map of the distortion vector, the correction vector components are simply pulled from the data-frame. While the projected lattice sites allow for easy atom column indexing, they are not the true undistorted lattice sites, as each iterative step accounts for previous distortions. To account for the accumulative nature of the indexing propagation, the ideal lattice with basis vectors $v_1$ and $v_2$ is also formed and correlated to the same $[i,j]$ index as the previous projected lattice. Now each atom column, projected lattice site, and ideal lattice site are coupled to a single $[i,j]$ index.

The ideal lattice constructed in a previous step, however, is not truly ideal. Prior to constructing a map of the atomic displacements from their ideal positions, these positions themselves must be refined; if the ideal lattice is skewed by even a fraction of a degree or pixel relative to the ground truth, artifacts such as Moiré patterns can appear. To overcome this issue, the position difference for each atom between the initial guess of the ideal lattice and the indexed real system is calculated and summed for all the atoms. This sum is minimized by employing an iterative technique that performs small corrections to the spacing and rotation of the ideal lattice. The Nelder-Mead simplex algorithm[36] was used to perform this minimization. Specifically, we minimized the summed total displacement, denoted by Z, which we define as:

$$Z = \sum_{i,j}(x_{i,j} - x_{i,j\ ideal})^2 + (y_{i,j} - y_{i,j\ ideal})^2 \qquad (3)$$

where the $(x_{ideal}, y_{ideal})$ coordinates are transformed by rotating and scaling the original ideal lattice. Upon minimization, we claim the ground truth ideal lattice is then obtained. The point estimate of the lattice parameters obtained in this way may have a systematic and uncorrectable error due to reliance on the acquired image - which is itself distorted - however the spatial structure of the distortion function will be extracted, since the ground truth ideal lattice is exactly periodic while the image is not perfectly periodic. The positions of this ground truth lattice are then compared to the positions in the real system and a displacement map at each atom location is calculated, which represents a sampling of the entire distortion.

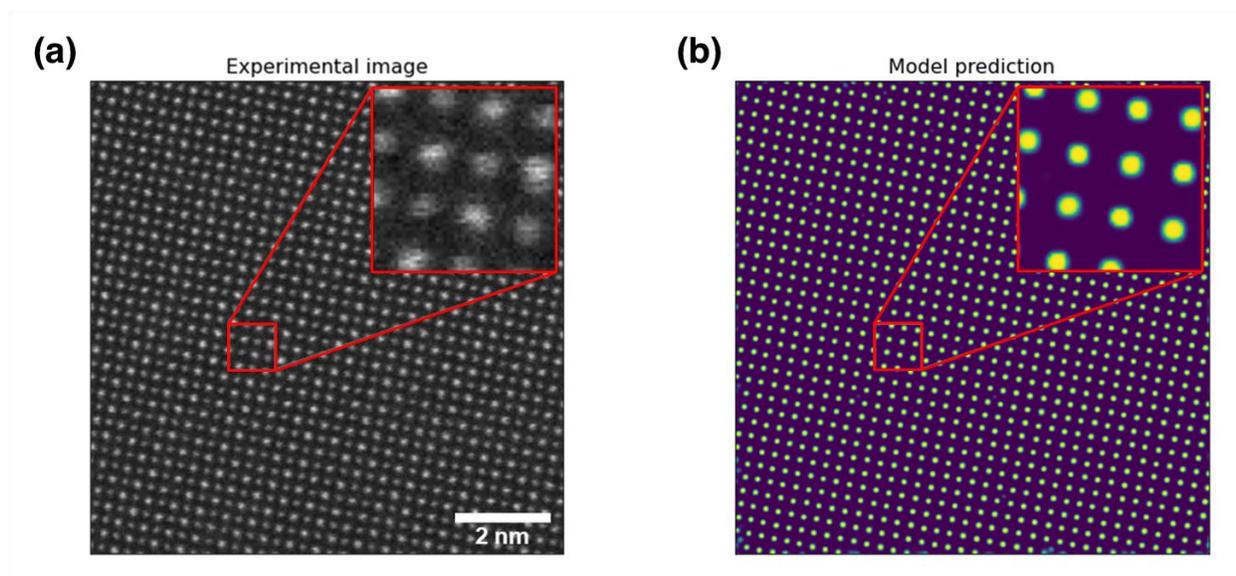

**Figure S1.** Atom finding on weak flyback data set. (a) shows the raw experimental image, while (b) shows the output of the neural network prediction. Note fewer atomic columns are present as compared to the strong flyback data set.

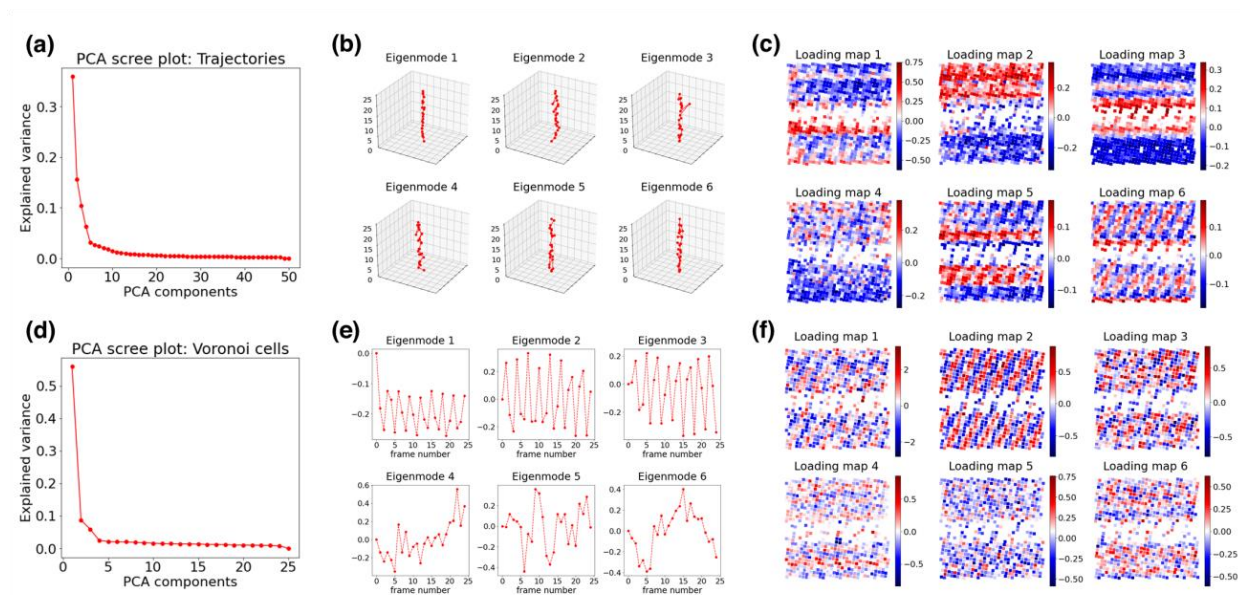

**Figure S2**. PCA analysis on trajectories (a-c) and Voronoi cell volumes (d-f) for a weak fly-back distorted image stack. Shown in each case are (a,d) scree plots, (b,e) eigenmodes, and (c,f) loading maps for the first six PCA components. Color scales in (c) and (f) are in units of pixels.

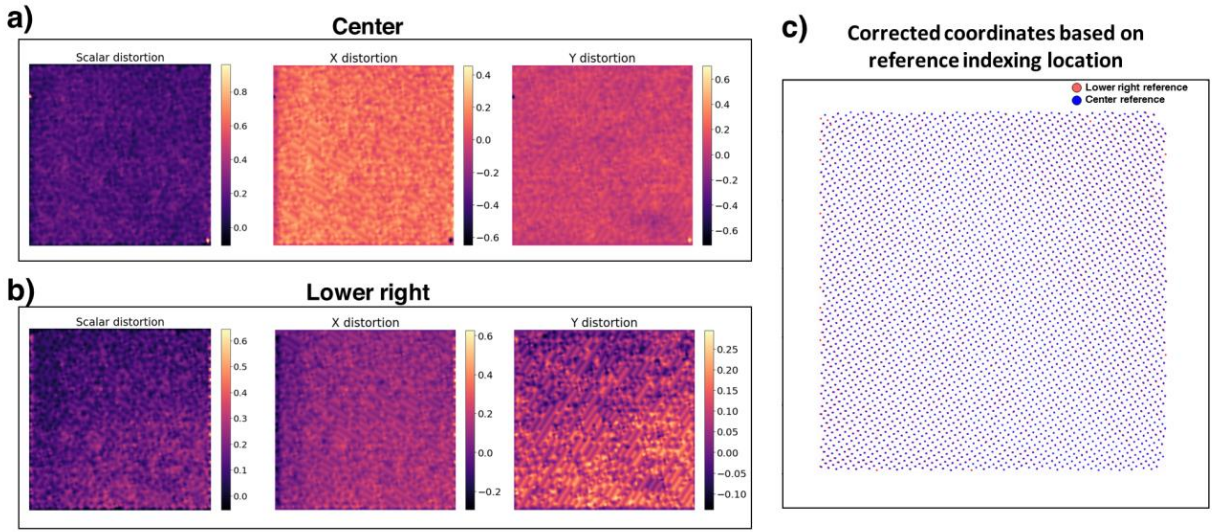

**Figure S3.** Effect of choosing different zero-distortion reference point. (*a*) shows correction using reference point in center of image while (*b*) shows correction using reference point in lower right quadrant of image. Overlay of corrected coordinates shown in (c) for comparison. Scale bars are in units of lattice spacings.

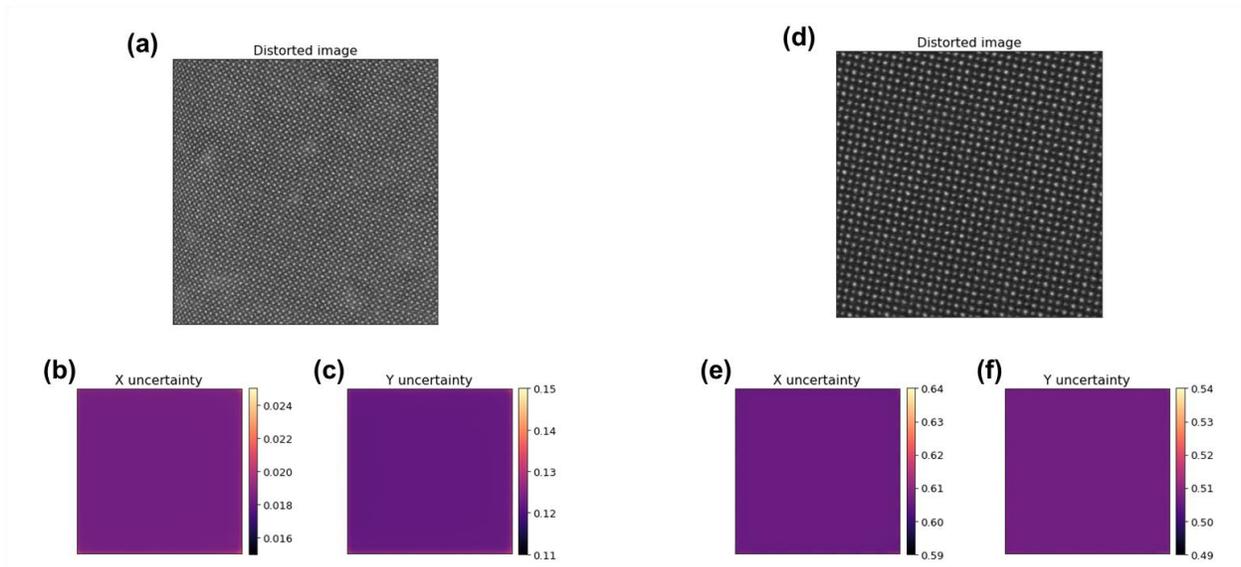

**Figure S4**. Uncertainty maps of *DF* obtained by GP method for long fly-back (a-c) and short fly-back (d-f) datasets. Note uncertainty is smooth across entire 2D space indicating the *DF* is well fit both on and off observation points (i.e., atom columns).